%% file: survey_main.tex
\documentclass[10pt,journal,cspaper,compsoc]{IEEEtran}
\usepackage{amsmath}
\usepackage{graphicx,epsfig,epstopdf,color,endnotes,alltt}
\usepackage{subfigure,caption}
\usepackage{listings}
\usepackage{txfonts}
\usepackage{url}
\usepackage{balance}
\usepackage{algorithm}
\usepackage[noend]{algorithmic}
\usepackage{fancyvrb}
\usepackage{bbding}
\usepackage[normalem]{ulem}
\usepackage{array}
\usepackage{multirow}
\usepackage{rotating,booktabs}
\definecolor{Magenta}{rgb}{1,0.5,0}

\newcommand{\amelie}[1]{{ #1}}
\begin{document}

\title{A Taxonomy and Survey on eScience as a Service
in the Cloud}

\author{Amelie~Chi~Zhou,
        Bingsheng~He
        and~Shadi~Ibrahim
\IEEEcompsocitemizethanks{\IEEEcompsocthanksitem A-C. Zhou and B-S. He, School of Computer Engineering,
Nanyang Technological University, Singapore. \protect\\
\IEEEcompsocthanksitem S. Ibrahim,
INRIA Rennes - Bretagne Atlantique, Rennes, France.}
\thanks{}}
\IEEEcompsoctitleabstractindextext{
\input{abstract}

}

%
%
%



\maketitle

\IEEEdisplaynotcompsoctitleabstractindextext
\IEEEpeerreviewmaketitle

\input{intro}
\input{background}
\input{taxonomy}
\input{status}
\input{discussion}
\input{conclusion}

\bibliographystyle{IEEEtran}
\bibliography{survey}

\end{document}

%% file: abstract.tex
\begin{abstract}
Cloud computing has recently evolved as a popular computing
infrastructure for many applications. Scientific computing, which
was mainly hosted in private clusters and grids, has started to
migrate development and deployment to the public cloud environment.
eScience as a service becomes an emerging and promising direction
for science computing. We review recent efforts in developing and
deploying scientific computing applications in the cloud. In
particular, we introduce a taxonomy specifically designed for
scientific computing in the cloud, and further review the taxonomy
with four major kinds of science applications, including life
sciences, physics sciences, social and humanities sciences, and
climate and earth sciences. Our major finding is that, despite
existing efforts in developing cloud-based eScience, eScience still
has a long way to go to fully unlock the power of cloud computing
paradigm. Therefore, we present the challenges and opportunities in
the future development of cloud-based eScience services, and call
for collaborations and innovations from both the scientific and
computer system communities to address those challenges.
\end{abstract}


%% file: intro.tex
\section{Introduction}\label{sec:intro}

The development of computer science and technology widens our view
to the world. As a result, the amount of data observed from the
world to be stored and processed has also become larger. Analysis of
such large-scale data with traditional technologies would be too time
consuming to hinder the development of scientific discoveries and theories.
eScience is the kind of science specifically proposed to address
large-scale data problems. It is the tool that offers scientists the
scope to store, interpret, analyze and distribute their data to other
research groups. eScience will play a significant role in every aspect of
scientific research, starting from the initial theory-based research though
simulations, systematical testing and verification to
the organized collecting, processing and interpretation of scientific data.
Recently, cloud
computing has been considered as the computing infrastructure for
eScience. This survey paper reviews the status of cloud-based
eScience and further identifies the challenges and opportunities
along this line of research.

Although the term of eScience has only been used for about a
decade, the study of eScience problems started much earlier. In the
early days, scientists from various fields couldn't really capture,
organize and analyze the large-scale scientific data, hindering the development of science. Technological advances
such as the computer and Internet have brought eScience study to
a new stage. eScience projects in various fields such as biology,
chemistry, physics and sociology are
emerging~\cite{A-brain-Antoniu,Hu:2010:BLS:1931470.1931831,Ekanayake:2008:MDI:1488725.1488926,Matsunaga:2008:CCM:1488725.1488913}, benefiting from the platforms and toolkits in computer science and
development experience shared by other research groups in domain
fields. Grid computing has greatly advanced the development of
eScience. Currently, almost all major eScience projects are hosted
in the grid or cluster environments~\cite{GridPP:2006}. With
aggregated computational power and storage capacity, grids are able
to host the vast amount of data generated by eScience applications
and efficiently conduct data analysis. This has enabled
researchers to collaboratively work with other professionals around
the world and to handle data enormously larger in size than before.
Many countries have devoted much investment to
build their own grid platform, such as GridPP~\cite{GridPP:2006} in
the UK and TeraGrid in US, CNGrid in China, and so on.

In the last few years, the emergence of cloud computing has brought
the development of eScience to another new stage. Cloud computing
has the advantages of scalability, high capacity and easy
accessibility compared to grids. Recently, many eScience projects
from various research areas have been shifting to cloud
platforms~\cite{li:escience,li:fault,humphrey:assessing}. eScience
as a service becomes an emerging and promising direction for science
computing. This survey focuses on the cloud services and techniques
adopted in current eScience projects from the infrastructure,
ownership, application, processing tools, storage, security, service
models and collaboration aspects.

The service model and well-developed tools in the cloud platform
have offered great opportunities for eScience research. The service
model of the cloud relieves the users from the low-level
infrastructure problems. Cloud resources are easy accessible, which
makes it possible for researchers in small organizations to deal
with large-scale data. The well-developed tools in the cloud,
including workflow systems such as DAGMan~\cite{dagman} and new
cloud oriented programming models such as MapReduce and DryadLINQ
greatly reduce the development cycle of the eScience projects and
the risk of development faults as well. People from database
community are building scientific databases such as SciDB to better
fit the requirements of eScience. Various experiments with eScience
projects conducted on both the cloud and clusters are revealing the
benefits of doing science on the cloud, helping researchers to make
their choices.

While offering new development opportunities for eScience, the cloud
platform also introduces new challenges for developing eScience
services. Due to the pay-as-you-go pricing model, users of the cloud
need to properly plan their execution, as it is not trivial to
minimize the cost. Furthermore, the easy accessibility and resource
sharing mechanism of cloud computing introduces security issues
around storing sensitive data in the cloud. In order to ensure the
confidentiality of their data from other cloud users, they need to
design their own security mechanism and implement them on the cloud.
A cloud platform also has the problem of data lock-in, because the
current cloud providers do not have standardization on the services
they provide. Thus, moving data from one cloud to another is not
trivial. All these challenges require hard work and close
collaboration between domain experts in computer science and
eScience.

Although previous work has surveyed eScience and cloud computing
separately, few of them have provided a review from the point of
view of eScience in the
cloud~\cite{Rimal:2009:TSC:1683301.1684085,escience:survey:web}.
Both eScience and cloud computing are rapidly developing and
becoming more mature. It is timely to examine the efforts and future
work for scientific computing in the cloud. This article focuses
especially on eScience projects in the cloud and comparing the
advantages and weaknesses with eScience in the grid to discuss the
obstacles and opportunities of eScience services in the cloud.

The rest of the article is structured as follows.
Section~\ref{sec:backgr} introduces the background information of
eScience history, grid-based eScience and cloud-based eScience.
Section~\ref{sec:taxo} gives the taxonomy of eScience in the cloud.
Section~\ref{sec:class} presents some eScience example projects
on the cloud from four different scientific
areas. Section~\ref{sec:discussion} discusses about the obstacles
and opportunities for eScience projects on the cloud. Finally,
Section~\ref{sec:concl} draws conclusions from the article.
\begin{figure}
    \centering
    \includegraphics[width=0.9\linewidth]{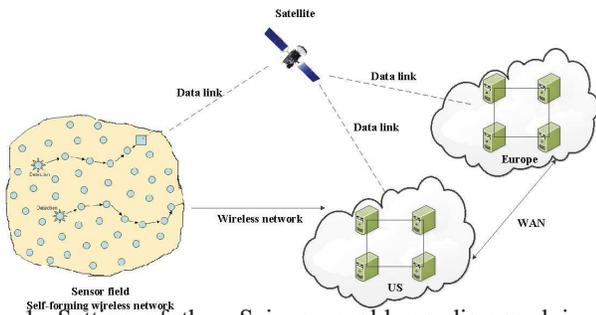}\vspace{-2ex}
    \caption{Setting of the eScience problems discussed in this article} \label{fig:intention}\vspace{-2ex}
\end{figure}

%% file: background.tex
\section{Background}\label{sec:backgr}

In this section, we briefly discuss some history remarks on
scientific computing development, particularly for eScience. Next,
we focus our review on the grid based scientific computing, and
introduce the background for cloud computing.

\subsection{History Remarks}

\begin{table}
\centering
\caption{Development stages of the scientific computing.\label{tb:stage}}{
    \begin{tabular}{ | p{1.5cm} | p{2cm} | c | c |}
    \hline
    Stage & Data Generated & Research Period & Infor. Tech.  \\ \hline
    Manual & By hand & Ad-hoc & Paper and pencil \\ \hline
    (Semi-) Automated & With the help of machinery & Short-term & Computer assisted \\ \hline
    Large-scale Sensing & From satellites and sensors around the world & Real-time & Cluster and grid  \\
    \hline
    \end{tabular}
    }\vspace{-2ex}
\end{table}

Due to intensive computational and data requirements from
scientific computing, computer infrastructures have been adopted to
host scientific data sets and computations. eScience is a new
science paradigm that uses distributed computing infrastructures for
computation, simulation and analysis. In addition, the scientists
can make use of high speed network to access huge distributed and
shared data collected by sensors or stored in database. This
distributed HPC and data environment allows scientists around the
world to share knowledge and resources, and build close scientific
collaborations.

The term \emph{eScience} was first proposed in 1999 and was further interpreted by more researchers since
then~\cite{defineescience}. During the development of eScience, we
believe it has gone through several stages to evolve from
traditional science to the eScience today. Table~\ref{tb:stage}
shows the major development stages the scientific computing has gone
through. We review the history in the following dimensions.

\emph{Dimension 1: the evolution of science.} We observed that
technology (particularly information technology) is one of the main
driving factors in pushing science forward. From the perspective of
experimental methods, eScience first used~\emph{manual} measurements:
meaning the measurements were taken by hand, not using machinery or
electronics to fulfill the function. Then with the development of
technology, machinery such as computers and metering instruments are
used to help in the measurements, but with manual operations still
involved. This stage is called the~\emph{semi-automated} stage.
After this stage, machinery took a greater part in the measurements
and eScience has evolved to the~\emph{automated} stage where
machines took almost all the work with the least of human
involvement. To recent years, new technologies such as high
performance computers, sensor networks and various experimental
softwares make the eScience measurements evolve to
the~\emph{large-scale sensing} stage~\cite{Globalmodelling}. Take
the research in Meteorology for example, in the early stage
(classified to manual stage), researchers use thermometer,
barometer, hygrometer and etc to measure the meteorological
variables such as temperature, air pressure, water vapor and write
down the records. They archive those meteorological data for drawing
climatic maps and studying the climate of local area.

In the 19th century (classified to semi-automated stage),
breakthroughs occur in meteorology after observing the development
of networks. The meteorological data collected in local
meteorological observatories are transmitted through networks and
then are gathered together by different spatial scales to study the
various meteorological phenomena.

Since the 20th century (classified to automated stage), with the
adoption of radars, lasers, remote sensors and satellites into the
meteorological research, collecting data of a large area is no
longer a challenging problem and special instruments together with
the automation of computers can automatically fulfill the measuring
tasks. During this time, computers are used for doing data analysis
and transmitting results for sharing. At the end of the 20th century
(classified to large-scale sensing stage), large scale observation
experiments are performed. Such as during December 1977 to November
1979, back then a large scale atmospheric measurement experiment
took place involving more than 100 countries around the world. This
experiment was relied on satellites, meteorological rockets,
meteorological observatories on the ground around the world,
automatic meteorological stations, airplanes, ships, buoy stations
and constant level balloons. These instruments were combined to form
a complete observing system to automatically measure the
meteorological variables world-wide.

\emph{Dimension 2: the length of research period.} eScience has gone
through \emph{ad--hoc} stage when research was done just for a
specific problem or task, and not for other general
purposes later; \emph{short-term plan}
stage when researchers made plans in priori for their problems about
what to do in what time, so that a project of a short term could be
kept on schedule; and \emph{real-time} stage when the research is
subject to real-time constraints, such as the experimental data are
collected in real-time and the system needs to give out results also
in real-time. This evolution on research period also require the
experimental methods to be more efficient, and the support of high
technology as we will discuss next.

\emph{Dimension 3: the technology.} eScience has gone through
\emph{paper and pencil} stage when no machinery was involved in our
research and human work with paper and pencil was the only tool for
science; then computers appeared and eScience was thus able to move
to the \emph{computer assisted} stage when computers played a great
role in helping with complex calculations and solving logical
problems; with the scientific problems getting more complicated and
traditional computers not sufficient for the computing power
required, \emph{cluster and grid} are coming to scientists' vision
and help them solving many data-intensive or compute-intensive
problems within reasonable time which is not possible on traditional
computers.

We summarize our findings in the three dimensions. Scientists only
deal with specific problems using manual methods such as doing
theoretical calculation using paper and pencil at early days. As
problems getting more complicated, more planning is needed for the
research and semi-automated and automated methods are also required in
the research during this time. Computers are used and when problem
scale gets larger, new technologies such as clusters and grids are
applied for solving the problems faster. What's the next step? When
problem scales get even larger and the big data coming into sight,
also with the real-time constraints on the problems, even clusters
and grids are not enough to tackle such problems. Recently, many
eScience projects are leveraging the technology of cloud
computing~\cite{A-brain-Antoniu,Matsunaga:2008:CCM:1488725.1488913,Newman:2008:SSS:1488725.1488885,Watson08cloudcomputing}.
With its high performance, scalable and easy accessible
characteristics, it will offer new opportunities for the new
problems.

\subsection{Grid-based eScience}

%

Current major eScience projects are mostly hosted in the grid or \amelie{HPC}
cluster environment. With aggregated computational power and storage
capacity, grids have been considered the ideal candidate for
scientific computing. There are many labs around the world working
on grid based projects, such as GridPP in UK, TeraGrid in US, CNGrid
in China, France Grilles in France, D-Grid in Germany, Kidney-Grid
in Australia, etc.

In UK, particle physicists and computer scientists have been collaboratively working
on the GridPP project. They manage and maintain a distributed computing grid across the UK with the primary aim of providing resources to particle physicists working on the Large Hadron Collider experiments at CERN~\cite{GridPP:2006}. The collaboration incorporates computing facilities at the Rutherford Appleton Laboratory along with four other grid organizations of ScotGrid, NorthGrid, SouthGrid and LondonGrid.
These organizations include all of the UK universities and institutions that are working as members of this project. At the end of 2011, the project has contributed a large number of resources (29,000 CPUs and 25 Petabytes of storage) to the worldwide grid infrastructure.

The Grid Infrastructure Group (GIG) along with eleven
resource provider sites in the United States have initiated
an eScience grid computing project called TeraGrid.
TeraGrid provides high-performance computation
resources, data resources and tools, and high-end experimental facilities to users all around the USA through high-performance network connections.
For example, in 2007, the resources TeraGrid provided included more than 250 Teraflops
of computation resources and more than 30 Petabytes of data storage resources. Researchers could access
more than 100 databases of different disciplines. In late 2009,
TeraGrid resources had grown to 2 Petaflops of computing capability
and more than 60 Petabytes storage. In mid-2009, US National Science
Foundation (NSF) extended the operation of TeraGrid to 2011.

China National Grid (CNGrid) has quickly grown to serve more than
1400 users including both research institutes and commercial companies,
providing more than 380 Teraflops of computation resources and more than 2 Petabytes of
shared data storage resources.
Since 2009, this project has built three Petaflop-level supercomputers, in which
Tianhe-1 was ranked the fastest supercomputer in the top 500 supercomputers in 2010~\cite{top500:2010}.
With the built of the three supercomputers, CNGrid resources has grown to 8
Petaflops of computation capability and supports computation services for
more than 700 national research and engineering projects in the areas of
meteorology, medicine and pharmacology, aircraft engineering and
aerospace engineering, etc.

Another example is the Worldwide LHC Computing Grid, which involves international
collaborations of more than 150 computing centers in nearly 40 countries around the world.
The European Grid Infrastructure (EGI)~\cite{egi},
the Open Science Grid and the Nordic Data Grid Facility, etc, are all participants of this project.
It consists of a grid-based computer network infrastructure to utilize the global computation
resources for storing, distributing and processing the large volume of data (around 25 Petabytes per year) produced by
the Large Hadron Collider (LHC) experiments. At the end of 2010, the Grid consisted of
200 thousand processing cores and 150 Petabytes of disk space, distributed across 34 countries.

\amelie{Besides the collaborations between major research centers, 
volunteer computing projects are taken place to build grid platforms with public donation of computing resources. SETI@home~\cite{seti:website} is such a volunteer computing project employing the BOINC software platform to search for
extraterrestrial signals with the spare capacity on home and office computers.}

The strength of grid computing has attract many scientific applications to work on grids.
\begin{itemize}
\item

First, since governments are very concerned about the research on
grid and frontier scientific research, most of the grid-based
projects are funded by national fundings. Such as the GridPP project
is funded by the UK's Science and Technology Facilities Council with
a total amount of 47 million pounds till 2011; the TeraGrid project
received 98 million dollars from NSF by 2004, 150 million dollars
extended support in 2005 and another 121 million in 2011; the CNGrid
project received around 94 billion Chinese yuan from 2006 to 2010. Sufficient
amount of money offers good chances for institutes to hire highly
qualified domain experts to do research and equip powerful computers
and other resources.
\item

Second, single research institute can enjoy the vast computational
and storage resources from grids by donating their own idle
resources. Such institutes may not have enough budget for them to
buy powerful computers or build their own data centers. 
\item

Third, the tools and softwares developed on grid can benefit more
research groups besides the developers themselves. This strength can
save a lot of development time for the projects developed on the
grids.
\end{itemize}

While Grid is the dominant infrastructure for eScience, it faces a
number of limitations. First, due to the development of sensors and
storage techniques, many data-intensive eScience applications are
emerging. Even with the powerful supercomputers, grid may no longer
satisfy the need of capacity. Second, due to the limitation of its
structure, grid is not able to provide the elasticity required by
most scientific projects which are pursuing cost efficiency. Third,
it's not easy to get access to grid resources for everyone because a
program getting access to grid resources needs to be authorized on
the project's behalf and resources would then be distributed to this
project as a whole. Since grids are mostly national-wide
initiatives, getting the authorization is very hard for most
small-scale projects. Finally, while Grid offers access to many
heterogeneous resources, many applications need very specific and
consistent environments. Due to these reasons, many of the eScience
applications are shitting to the Cloud which has elastic storage and
computing power.

\subsection{Cloud Computing}

According to the definition of the National Institute of Science and Technology (NIST), cloud computing is\\

\emph{``The delivery of computing as a service rather than a product, whereby shared resources, software, and information are provided to computers and other devices as a utility (like the electricity grid) over a network (typically the Internet)''~\cite{clouddefinition}.\\}

Cloud computing hasn't come into popularity until early 2000's, when a lot of research efforts on the cloud were emerging.
Officially launched in 2006, Amazon Web Service (AWS) is the first utility computing platform that provides computation resources as services to external customers. Many other cloud service providers, including Microsoft (Microsoft's Azure), Google (Google's Cloud Platform) and OpenStack, have come into the market since then.
Open-source systems and research projects are developed to facilitate the use of cloud. Initially released in early 2008, Eucalyptus is an open-source system for deploying AWS-compatible private and hybrid cloud computing environments. In the same year, the OpenNebula toolkit was released, which is also designed for building private and hybrid clouds but with different design principles from Eucalyptus.

Cloud computing bares many similarities and differences with grid
computing. In the year 2008, Foster et al.~\cite{CloudGrid360} has
compared clouds and grids mainly from a technological perspective.
Five years have passed, and we should take a revisit on those
differences to catch up the recent rapid development of cloud
computing and highlight its relevance to the requirement of
eScience.

Compared to the grid, cloud has better scalability and elasticity.
\begin{itemize}
\item When developing applications on the grid infrastructure, it's not easy to scale up or down according to the change of data scale. But in cloud, with the use of virtualization, clients can scale up or down as they need and pay only for the resources they used.
\item Virtualization techniques also increase the computation efficiency as multiple applications can be run on the same server, increase application availability since virtualization allows quick recovery from unplanned outages with no interruption in service and improves responsiveness using automated resource provisioning, monitoring and maintenance.
\item Also, cloud has easier accessibility compared to grid. Users can access to commercial cloud resources through log in and use the resources as they need as long as they could pay with a credit card. In this case, even small businesses which could not afford purchasing high performance computers can also have the chance to use powerful clusters or supercomputers on their compute-intensive or data-intensive projects.
\end{itemize}
eScience applications are beginning to shift from grid to cloud
platforms. The Berkeley Water Center is undertaking a series of
eScience projects collaborating with
Microsoft~\cite{li:escience,li:fault,humphrey:assessing}. They
utilized the Windows Azure cloud to enable rapid scientific data
browsing for availability and applicability and enable environmental
science via data synthesis from multiple sources. Their BWC Data
Server project is developing an advanced data synthesis server.
Through close interaction between computer scientists and
environmental scientists, they are building new tools and approaches
to benefit regional and global scale data analysis
efforts~\cite{li:escience,li:fault}. Another one, the California
Water CyberInfrastructure project, is developing a Water
Cyberinfrastructure prototype that can be used to investigate and
eventually manage water resources. In Europe, GRNET is initiating an
eScience cloud in Greece~\cite{grnet:greek}. GRNET is a state-owned
company operating under the supervision of the Ministry of Education
(General Secretariat of Research \& Technology). Its main mission is
to provide high-quality electronic infrastructure services to the
Greek academic and research institutions. The vision of its eScience
cloud is to provide virtualization and storage services for the
Greek scientific community. This project starts with offering online
storage of 50Gbytes for all Greek academic and research community
(Pithos), then moves to provide VMs on demand and finally provide
software as a service. We are also working on a eScience project
based on cloud computing in Singapore~\cite{water:pdcc}. The
objective of our project is to leverage cloud computing techniques
and sensor networks to provide real-time and large-scale monitoring
and analysis for water quality. The project is aiming at providing
real-time monitoring for the reservoirs based in Singapore, but the
methods and models proposed could be utilized to benefit all water
resources around the world. This project is funded by NRF Singapore
and we are currently working on the first phase.

Since the cloud is an emerging field, many of the cloud based
eScience projects are still in their early stage. This is partially
because cloud computing has come to popularity only for several
years and researchers haven't realized its strengths thoroughly.
That motivates us to review the existing efforts on adopting cloud
computing technologies to eScience, and to explore the research
challenges and opportunities in that direction. Moreover, a taxonomy
is useful in guiding the design and implementation of cloud-based
eScience project.

Compared with the general cloud computing surveys (such as by
Armbrust et al.~\cite{Armbrust:2010:VCC:1721654.1721672}), this
survey focuses on the review on the current status of eScience in
the cloud, and therefore identifies the new opportunities and
challenges on pushing the state-of-the-art. Our survey also goes
beyond some perspective report on science cloud (for example, by Lee
~\cite{Lee:2010:PSC:1851476.1851542} and by
Keahey~\cite{keahey:cloud} and Oliveira~\cite{Oliveira2010} in three
major aspects. First, we define a taxonomy for eScience services in
the cloud. To the best of our knowledge, our definition is the first
taxonomy for eScience services. Second, we perform the detailed and
comparative study on the existing efforts including tools, systems
and projects. Second, based on the review on the existing efforts,
we point out the challenges and opportunities that are close and
practical as a guide for the intermediate next steps.

%% file: taxonomy.tex
\section{Taxonomy of eScience Services in the Cloud}\label{sec:taxo}

\begin{figure}
    \centering{
  \includegraphics[width=0.9\linewidth]{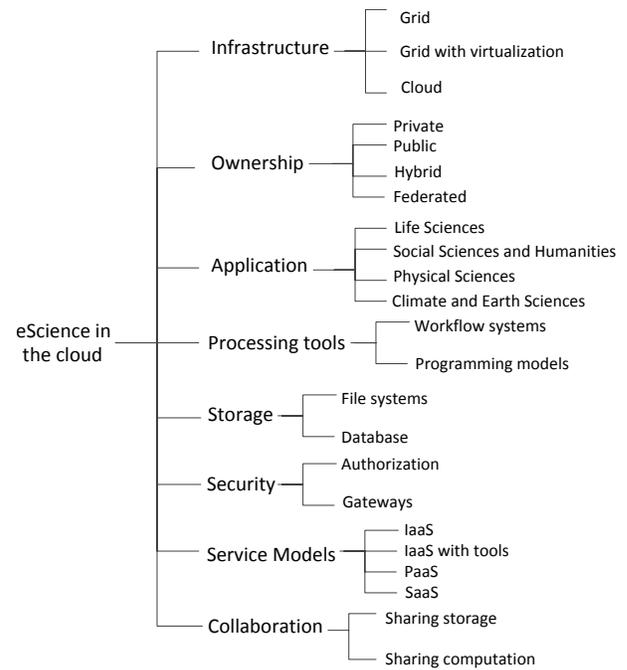}
    \caption{Taxonomy of eScience in the Cloud} \label{fig:taxonomy}}\vspace{-2ex}
\end{figure}

The taxonomy in this section gives clear classification of cloud
computing techniques used in eScience services from various
perspectives, including the computation \emph{infrastructure} for
eScience applications, the \emph{ownership} of cloud
infrastructures, the eScience \emph{application} types, the
\emph{processing tools} used for eScience applications, the
\emph{storage} model, the \emph{security} insurance method,
\emph{service models} of the cloud and the \emph{collaboration} goal
between different research groups. Figure~\ref{fig:taxonomy} gives a
clear structure of the taxonomy. This taxonomy reflects the
interplay between eScience and cloud computing. Some are mainly from
eScience's perspective, and some are mainly from cloud computing's
perspective. We introduce them one by one.

\vspace{-1.5ex}
\subsection{Infrastructure}
The infrastructure of cloud provides access to compute and storage
resources for eScience applications in an on-demand fashion. Cloud
shares some similarities with Grid while at the same time is
modified to overcome the limitations of Grid.

Grid computing technologies are leveraged by cloud computing to
serve as its backbone and infrastructure support. Compared with
grids infrastructures, cloud has pricing and monitoring services.
Before 2007, most of eScience applications were implemented on Grid,
where scientific organizations share their spare resources.

One characteristic of Grid is that it assigns resources to users in
the unit of organizations and each individual organization holds
full control of the resources assigned to it. However, this kind of
resource assignment is not efficient. There are efforts in Grid to
use virtualization to change this situation. Nimbus scientific cloud
is one such effort that provides a virtual workspace for dynamic and
secure deployment in the Grid.~\cite{Hoffa:2008:UCC:1488725.1488955}
is an astronomy application implemented using Nimbus. Virtualization
hides from users the underlying infrastructures which are usually
heterogeneous hardware and software resources, and provides the
users with homogeneous and isolated virtual cloud environment.

In contrast to science clouds, several national cloud initiatives have
also been announced to provide on-demand resources for governmental
purposes~\cite{Lee:2010:PSC:1851476.1851542}, such as the US Cloud
Storefront~\cite{us:store}, the UK G-Cloud~\cite{uk:gcloud}, and the
Japanese Kasumigaseki~\cite{Kasumigaseki} cloud initiatives. Many
industry players also dive in the cloud business and provide users
with seemingly infinite public cloud resources. With the popularity
of cloud, many eScience applications are right now transferring to
the general public cloud infrastructures such as Amazon EC2, Windows
Azure to benefit from its high performance, scalability and
easy-access~\cite{li:escience,li:fault,humphrey:assessing,Subramanian:2010:RPS:1931470.1931899,citeulike:3523379,Nunez:2010:NMP:1932688.1933014}.

\vspace{-1.5ex}
\subsection{Ownership}
The ownership of cloud infrastructures can be classified as the following types: private, public, hybrid and federated.

Private clouds are infrastructures operated only for one single
organization, no matter who the infrastructures are managed by or
where they're located. The security level of private clouds is the
highest among the four types. eScience applications which have high
security requirements or posses highly sensitive data can be
deployed on private clouds. OpenNebula is the first open-source
software supporting private clouds deployment and is widely used by
industry and research users right now~\cite{opennebula}. But on the
other hand, such infrastructures do not benefit from the economic
models provided by the cloud since the application owners have to
``buy, build and manage'' the infrastructures to run their jobs.

In contrary, public clouds are more open, with their application,
storage and other resources available to the public on the
pay-as-you-go basis. There are quite a few commercial companies
providing public cloud services, such as Amazon, Windows and Google.
Many eScience applications have been deployed on this kind of cloud
platforms (e.g.,~\cite{citeulike:3523379,li:escience,li:fault})
because users can easily access to the public cloud resources with a
credit card.

\amelie{A federated cloud, also known as community cloud, is a
combination of two or more clouds from either
private, public or federated clouds.} In this combination,
the two or more clouds often have
common goals in security, compliance, jurisdiction, etc. Many
countries have built federated clouds to support the research and
education purpose of their own country. The EGI Federated Cloud Task
Force~\cite{federatedtf} is a federation of academic private clouds
to provide services for the scientific community. It has been used
by a wide areas of eScience applications, including Gaia which is a
global space astrometry mission~\cite{gaiaspace}, the Catania
Science Gateway Framework (CSGF)~\cite{catania} which provides
science gateway for scientific application users, etc.

\amelie{A hybrid cloud utilizes cloud resources from both private and public clouds. }
The benefit of hybrid clouds
captures the best of both worlds. When the resources of the private
cloud are enough to support current workload, the users will only
use the private cloud to benefit from its security and stability.
While the workload is bursting and the private cloud can no longer
support users' requirements, users can then request resources from
the public cloud to benefit from its scalability.

\vspace{-1.5ex}
\subsection{Application}
Cloud computing techniques have been applied to various eScience
applications. We have surveyed a lot of eScience papers and
summarized them in the following four categories based on their
areas of expertise: Life
sciences~\cite{Matsunaga:2008:CCM:1488725.1488913,Newman:2008:SSS:1488725.1488885},
Physical
sciences~\cite{Deelman:2008:CDS:1413370.1413421,Hoffa:2008:UCC:1488725.1488955},
Climate and Earth sciences~\cite{citeulike:3523379,li:escience} as
well as Social sciences and
Humanities~\cite{Curry:2008:FMV:1437901.1438789,Markatchev:2009:CIA:1723206.1724809}.

The life sciences comprise the scientific research on
living organisms, such as plants, animals, and
human beings. Specifically, it includes the fields of Biochemistry,
Biology, Ecology, Neuroscience, Psycology, etc. Physical sciences
encompass the fields of natural science and science that study
non-living systems, in contrast to the life sciences. Climate and
Earth science is the study of climate and the planet Earth. The climate
science is a sub-field under the atmospheric sciences which studies the
average weather conditions in a period of time while the earth science
includes the study of the atmosphere, hydrosphere, oceans and biosphere,
 as well as the solid earth. Social sciences and Humanities is the field of study
concerned with society and human behaviors. It includes the scientific studies
on anthropology, archeology, criminology, economics, education, linguistics,
political science and international relations, sociology, geography,
history, law, and psychology.

We note that those application categories can have overlaps with
each other. There is no absolute boundary between each pair of
categories. Still, different categories have their own requirements
on the cloud. The first three categories, life sciences, physical
sciences and climate and earth sciences, are more focusing on
extending their works to large-scale datasets and thus require the
cloud platform to deal with large-scale data analysis efficiently.
The fourth category, social science and humanities, is more focusing
on collaboration and thus
requires the cloud platform to be easy for sharing. 

\vspace{-1.5ex}
\subsection{Processing tools}
From the perspective of processing tools, we have witnessed deployment of classic workflow systems in the cloud, new cloud oriented programming models such as MapReduce and DryadLINQ, and hybrid of such newly proposed tools and models.

Scientific workflows have been proposed and developed to assist
scientists to track the evolution of their data and results. Many
scientific applications use workflow systems to enable the
composition and execution of complex analysis on distributed
resources~\cite{Deelman08workflowsand}. Montage is the example of a
widely used workflow for making large-scale, science-grade images
for astronomical research~\cite{Hoffa:2008:UCC:1488725.1488955}.

\amelie{Workflow management systems (WMSes) such as Pegasus~\cite{Vockler:2011:EUC:1996109.1996114} and Kepler~\cite{Wang20121630} are developed to manage and schedule the
execution of scientific workflows. WMSes rely on tools such as Directed Acyclic Graph Manager (DAGMan)~\cite{dagman} and Condor~\cite{condor} to manage the task dependencies within scientific
workflows, and to manage the resource acquisition from the cloud and
schedule the tasks of scientific workflows to cloud resources for execution.}

The appearance of cloud oriented programming models has great
promotion for the development of cloud computing. MapReduce is a
framework proposed by Google in 2004~\cite{Dean:2008:MSD:1327452.1327492}
for processing highly distributable problems using a large number of computers.
The framework is inspired
by the map and reduce functional language where the map function
takes in the input, partitions it into smaller sub-problems and
distributes them to multiple worker machines while the reduce
function collects the processing results to all the sub-problems and combines
them in some way to form the output. Users who need to parallel
their codes in order to run in distributed environment only need to
define their own map and reduce functions using the MapReduce
framework. This makes this framework especially suitable for
eScience application users who may not be experts in parallel
programming. We have observed the emergence of eScience applications
adopting MapReduce framework for data-intensive scientific
analyses~\cite{Ekanayake:2008:MDI:1488725.1488926}.

\vspace{-1.5ex}
\subsection{Storage}
Data is centric to eScience applications and data processing is
closely related to data storage. With the development of science,
the hypothesis to data has evolved from empirical description stage,
theoretical modelling stage, computational simulation stage to the
fourth paradigm today, the data-intensive scientific discovery
stage. Due to the vast data size, knowledge on the storage format of
scientific data in the cloud is very important. Normally, there are
two ways for data storage: data can be stored as files in file
systems or in databases.

Many distributed file systems have been proposed to provide
efficient and reliable access to large-scale data using
clusters of commodity
hardware~\cite{Ghemawat:2003:GFS:945445.945450,Shvachko:2010:HDF:1913798.1914427}.
For example, Hadoop Distributed File System (HDFS) is the primary
storage system used by Hadoop applications which utilize the
MapReduce model for large dataset processing. It creates multiple
replicas of data blocks and distributes them on compute nodes
throughout a cluster to enable reliable and rapid data access. When
well-designed, features of the HDFS system such as data locality and
data replication can further benefit the applications running on
Hadoop via locating computation close to the
data~\cite{Ramakrishnan:2010:DFP:1807128.1807145}. 
\amelie{OpenStack Swift~\cite{swift} is a distributed storage system for unstructured data at large scale. It currently serves the largest object storage clouds, such as Rackspace Cloud Files and IBM Sftlayer Cloud.} The scalable and
highly efficient distributed file system models provide a promising
data storage approach for data intensive eScience applications.

Although in cloud, data storage usually relies on file systems,
using databases as storage has its advantages. First, it's easier to
do query in a database than in file systems since the files have to
be opened and closed in order to get the data stored in. Also,
database as storage can ensure data integrity. Till now, the
parallel capabilities and the extensibility of relational database
systems (RDBMS) were successfully used in a number of
computationally-intensive analytical applications. When facing
eScience applications, RDBMS have shown limitations. For one thing,
not all data in eScience is relational. Several classes of ``NoSQL"
databases have been proposed as alternatives to RDBMS to satisfy the
efficiency requirement of scientific data. For example, Amazon's
Dynamo~\cite{DeCandia:2007:DAH:1294261.1294281} is a key-value store
which supports storing and retrieving data by primary key. Its
key-value interface makes it especially simple and cost-effective to
the cloud users. Google's
Bigtable~\cite{Chang:2006:BDS:1267308.1267323} is a column-oriented
NoSQL database which provides column-wise as well as row-wise index
for data manipulation. This distributed storage system is designed
to managing large-scale structured data: ``petabytes of data across thousands of commodity
servers''~(\cite{Chang:2006:BDS:1267308.1267323}).  
\amelie{Cassandra~\cite{Lakshman:2010:CDS:1773912.1773922} is another column-oriented distributed NoSQL database which provides highly available service to large amounts of structured data. }
HBase, a Hadoop
project modeled on Bigtable, has been applied to many eScience
applications such as bioinformatics
domains~\cite{citeulike:8467579}. Some array-based databases such as
SciDB~\cite{Brown:2010:OSL:1807167.1807271} have also been proposed
to satisfy the special requirement of array-based eScience
applications. SciDB is a scientific database system built from
ground up and has been applied to many scientific application areas,
including astronomy, earth remote sensing, environmental studies and
etc~\cite{scidb:usecase}.

\vspace{-1.5ex}
\subsection{Security}
Security is a big issue to eScience applications, especially for
those with sensitive data. On the one hand, scientists need to make
sure that the sensitive data is not stolen by people with vicious intension;
on the other hand, they also need to share data between scientific
groups working on the same project. Thus, how to find a balance
point between the two aims is a challenging problem. Currently, the
security level in the cloud is not very high compared to the Grid
computing platform and the common way to make sure of security in
the cloud is through logging in. Many eScience applications deployed
on the cloud have designed their own way of authentication and
authorization to ensure security. Such as
in~\cite{10.1109/eScience.2011.16}, Group Authorization Manager is
used to grant access permission based on user-defined access control
policy. The emerging Open Authorization (OAuth2.0) protocol is used
to support authorization for users to share datasets or computing
resources. In~\cite{Watson08cloudcomputing}, the Gold security
infrastructure is utilized to deal with the authentication and
authorization of users to keep sensitive data secure. Data owners
could specify their security preferences for the security
infrastructure to control role and task based access.

Unlike in Grid computing, where the authentication and authorization
mechanisms are mainly based on the public key infrastructure (PKI)
protocol~\cite{Alfieri05fromgridmap-file}, many Cloud vendors support
multiple security protocols such as OAuth2.0.
The adoption of the new security protocols opens up a new design space
for users to define rules of accessing secured resources and sharing data.
Via the authorization delegation in the security protocols,
users can define rules to allow easy collaborations between geographically
distributed parties without the involvement of administrators.

\vspace{-1.5ex}
\subsection{Service Models}
There are different levels of computing services offered by the cloud (i.e., IaaS, IaaS with tools, PaaS and SaaS). The IaaS model is the most basic cloud service model, where cloud providers only offer physical infrastructures to users, in the form of virtual machines, raw storage, and so on. Amazon EC2 is such an example~\cite{citeulike:3523379,Nunez:2010:NMP:1932688.1933014,eScience.2011.15}.
When deploying in an IaaS cloud, users have only to install operating system and application softwares as they need. In order to save users' effort of installation, platforms providing IaaS level services but with additional tools and softwares, have been proposed. Nimbus~\cite{Hoffa:2008:UCC:1488725.1488955} and Eucalyptus are examples of this kind. Nimbus is an open-source toolkit that aims to deliver IaaS
capabilities to the scientific community. It allows users to rapidly develop custom community-specific solutions.
In the PaaS model, cloud providers provide a computing platform typically equipped with operating system, programming language execution environment, database, and web server. Users of PaaS cloud can simply develop their applications on the platform without the effort and cost of buying and managing the underlying hardware and software layers. Typical examples of this type include Windows Azure, Google's App Engine.
In the SaaS model, cloud providers provide a computing platform installed with application softwares. Cloud
providers are in charge of the software maintenance and support. Cloud users are eased from the trouble of managing the cloud platform and can put more of their effort on application design. Notable service providers in this class include online storage services such as Dropbox and Google Drive, online education services such as Coursera.

\vspace{-1.5ex}
\subsection{Collaboration}
Another important usage of cloud for eScience applications is to
realize collaboration. In eScience, there are more and more projects
involving multiple groups closely working together on the same
project and those groups are sometimes spread worldwide. The
collaboration between the groups includes two different focuses. First
is on sharing storage, that is the sharing of scientific data and
analysis results between different research groups working on the
same project. Except sharing data for collaborative works, many
eScience applications open their data to the public for educational
purposes. Second is on sharing of computation, that is to share the
idle computing resources of one group to the others such that the
resource utilization rate of all the collaborating groups can be
highly improved. Collaboration between these groups is very
important to the success of the projects. With the development of
Internet and the popularity of social networks, many works are
leveraging cloud computing techniques and social network APIs to
provide a collaboration platform for eScience
researchers~\cite{Thaufeeg:2011:CES:2116259.2116588,10.1109/TSC.2011.39}.

%% file: status.tex
\section{Current Status}\label{sec:class}
We review the current status of eScience services in the cloud, and
present the key observations from our survey. 

\vspace{-1.5ex}
\subsection{An Overview}
The example systems surveyed in this section may not be exhaustive,
but cover many areas of eScience researches currently going on. The
table below summarized the systems from their platform, scientific
operations and development and classified them by their areas from
life sciences, social sciences and humanities, physical sciences and
climate and earth sciences. Table~\ref{tb:status} is a
categorization of the surveyed example systems using the taxonomies
introduced above. In the rest of this section, we present the major
observations we have found from the example systems.

\begin{sidewaystable*}\footnotesize
\centering
\caption{Taxonomy mapping to the surveyed example systems. ``-'' means that aspect is not specified in the project paper.\label{tb:status}}{
\tabcolsep=0.11cm
\begin{tabular}{ | l | p{1.5cm} | l | l | p{3cm} | p{1.5cm} | l | l | p{2cm} |}
\hline
Project&Infrastructure&Ownership&Application&Processing Tools&Storage&Security&Service Model&Collaboration\\ \hline
CloudBLAST&Grid with virtualization&Private&LS&MR Programming Model&File System&Authorization&IaaS&Computation\\ \hline
RDF&Cluster&Private&LS&MR Programming Model&Database&-&-&Computation\\ \hline
CARMEN&Cloud&Public&LS&Workflow System&File System / Database&Authorization&SaaS&Storage / computation\\ \hline
MFA&Cloud&Public&LS&Workflow System / MR Programming Model&File System&Authorization&IaaS&Computation\\ \hline
MassMatrix&Cloud&Public&LS&Workflow System&Database&-&IaaS with tools&Computation \\ \hline
LS Gateway&Cloud&Private&LS&Workflow System / MR Programming Model&File System&Gateway&SaaS&Storage / computation\\ \hline
CloudDRN&Cloud&Public&LS&Business Software Tools&Database&Authentication / Authorization&IaaS / PaaS & Storage \\ \hline
SciHmm&Cloud&Public&LS&Workflow System&Database&Authorization&IaaS&Computation\\ \hline
SciDim&Cloud&Public&LS&Workflow System&File System&-&IaaS&Computation\\ \hline
Montage Example&Cloud&Public&PS&Workflow System&File System&-&IaaS&Computation\\ \hline
Montage Comparison&Cloud&Private&PS&Workflow System&File System&-&IaaS with tools&Computation\\ \hline
CGL-MapReduce&Cluster&Private&PS&MR Programming Model&File System&-&-&Computation\\ \hline
Kepler&Grid / Cloud&Private / Public&PS&Workflow System&File System&-&IaaS&Computation\\ \hline
Inversion&Cloud&Public&PS&Programming Model&File System&Authorization&IaaS&Computation\\ \hline
CAOCM&Cloud&Public&CES&MPI Programming Model&File System&Authorization&IaaS&Computation\\ \hline
MODISAzure&Cloud&Public&CES&Workflow System&File System&Authorization&PaaS&Storage / computation\\ \hline
RPSS&Cloud&Public&CES&Multi-threading Programming Model&File System&-&PaaS&Storage / computation\\ \hline
NG-TEPHRA&Grid / Cloud&Private / Public&CES&Workflow System&File System&-&IaaS&Computation \\ \hline
Cloudbrusting&Grid / Cloud&Private / Public&CES&Workflow System&File System&-&PaaS&Computation\\ \hline
SLOSH&Cloud&Public&CES&Workflow System&File System&-&PaaS&Computation\\ \hline
FMVE&Cloud&Private&SSH&Programming Model&File System&Authorization&IaaS&Storage\\ \hline
IAS&Cloud&Public&SSH&-&File System&Gateway&SaaS&Storage / computation\\ \hline
BetterLife2.0&Cloud&Public&SSH&MR Programming Model&File System&Authorization&IaaS&Computation\\ \hline
SoCC&Cloud&Public&SSH&-&File System&Authorization&PaaS&Computation\\ \hline
SCC&Cloud&Public&SSH&-&File System&Authorization&SaaS&Storage / computation\\ \hline
\end{tabular}
}
\end{sidewaystable*}

\vspace{-1.5ex}
\subsection{Observation 1: Ad Hoc Project Developments}
The development of eScience projects is ad hoc. Some applications
are developed on Amazon EC2
cloud~\cite{Dalman:2010:MFA:1932688.1933004}, some are deployed on
Windows Azure~\cite{humphrey:assessing} while some others are
developed on both cloud platforms to verify their
design~\cite{conf/eScience/MudgeCHT11}. However, it is not clearly
explained why certain cloud platforms should be chosen over others in
those projects.

For example, MFA~\cite{Dalman:2010:MFA:1932688.1933004} is a Life Science project developed with the cloud services provided by Amazon. Its aim is to investigate whether utilizing MapReduce framework is beneficial to perform simulation tasks in the area of Systems Biology.
The Monte Carlo bootstrap (MCB) method, an important building block of this application, is parallelized and implemented with Amazon Elastic MapReduce (EMR). Because of the long-running characteristic of MCB simulation, the MapReduce version of MCB is wrapped with a WSRF service which is specifically designed to support long-running operations. The experiments on a 64 node Amazon MapReduce cluster and a single node
implementation have shown up to 14 times performance gain, with a total cost of on-demand resources of \$11.
MODISAzure~\cite{li:escience} is a Climate and Earth science application deployed on Windows Azure to process large scale satellite data. The system is implemented with the Azure blob storage for data repository and Azure queue services for task scheduling.
However, neither of the two projects has technically explained their choice of cloud platforms. The MapReduce framework is supported by many cloud providers other than Amazon, such as Cloudera's Distribution of Hadoop (CDH), Azure HDInsight, etc. The storage and queue services are also supported by many cloud providers besides Azure. For example, Amazon provides S3 and EBS for storage and Simple Queue Service (SQS) correspondingly.
To compare the performance on different cloud platforms, an Physical science project Inversion ~\cite{conf/eScience/MudgeCHT11} deployed its application on both Amazon EC2 and Windows Azure with symmetry structures.

All these examples indicate that, current eScience application
owners are not quite clear how to choose cloud platforms. The only
way for them to distinguish the cloud performance differences is
through redundant implementation. Due to the diverse requirements of
projects in different areas, the lessons learned during the
implementation of one project may be useless to projects in other
research areas.

\vspace{-1.5ex}
\subsection{Observation 2: Common Development Softwares and Tools}
Many eScience projects, especially those in the same application
class, share common development softwares and tools. For example,
a workflow system such as Pegasus is widely used by Physics science
applications~\cite{Deelman:2008:CDS:1413370.1413421,Hoffa:2008:UCC:1488725.1488955}
where the jobs involve a number of analysis steps. Many works are proposing new
techniques in the cloud for scientific applications based on the workflow model~\cite{Malawski:2012:CDP:2388996.2389026,DBLP:journals/pvldb/OgasawaraOVDPM11}. However, current
commercial clouds do not include such scientific tools by default.
Different applications owners have to redundantly deploy and

In the Life science project MassMatrix~\cite{eScience.2011.15}, the authors used the Pegasus Workflow Management System (WMS) to create parallel workflows for a database program which searches proteins and peptides from tandem mass spectrometry data. DAGMan is used to manage the data dependencies in the workflow and Condor is used to schedule the workflow. Similarly, the Physical science projects Montage Comparison~\cite{Hoffa:2008:UCC:1488725.1488955} and Kepler~\cite{Vockler:2011:EUC:1996109.1996114} also utilized Pegasus-WMS, DAGMan and Condor to manage the execution of astronomy workflows. In all three applications, the authors have to separately deploy and configure all the required softwares such as Pegasus and Condor on the cloud platforms to make their applications run. Such re-implementation and re-design work requires good effort from the application owners and should be avoided.

\vspace{-1.5ex}
\subsection{Observation 3: Static Data Storage}
Data is the centric of eScience applications. Although the data size
of most eScience applications is enormous, we have observed that many
of the eScience data are statically stored. For example, the
SciHmm~\cite{10.1109/eScience.2011.17} project is making
optimizations on time and money for the phylogenetic analysis problem.
The data involved in this application are genetic data, which do not
require frequent update and can be viewed as statically stored.
Similarly, the bioinformatics data in the
CloudBLAST~\cite{Matsunaga:2008:CCM:1488725.1488913} project and the
astronomy data in the Montage
Example~\cite{Deelman:2008:CDS:1413370.1413421}, although may be
updated from time to time, are seldomly modified once obtained. Once
such data are uploaded to the cloud, not much networking is required
to modify them. This characteristic of scientific applications makes
them appropriate to be implemented on the cloud since networking
usually causes the most monetary cost and overhead.

\vspace{-1.5ex}
\subsection{Observation 4: Privacy vs. Sharing}
Data privacy and security is a big issue to scientific applications.
Traditional storage at Grid and private clusters provides a high
security level to scientific data through authorization and
authentication. However, there is an increasing need of eScience
applications to collaborate and share. Such need forces them to move
their applications from traditional computing platforms to the
public cloud, which in turn makes the privacy issue more serious.

One example is the Life science project CloudDRN~\cite{clouddrn:escience2013}. CloudDRN moves medical research data to the cloud to enable secure collaboration and sharing of distributed data sets. It relies on authentication and authorization to ensure security. Also, many applications in Social Science and Humanities have shown such a trend. The SoCC~\cite{Thaufeeg:2011:CES:2116259.2116588} project leverages social network platform for the sharing of resources in scientific communities. They provide a PaaS social cloud framework for users to share resources and support creating virtual organizations and virtual clusters for collaborating users. The SCC~\cite{10.1109/TSC.2011.39} project is also leveraging social network and cloud computing to enable sharing between social network users. But different from previous works, it argues that since online relationships in social networks are often based on the real world relationships, it can be used to infer the trust levels between users. The benefit is users can thus share data and applications with lower privacy concerns and security overheads. In both examples, the social network information is utilized to lower the privacy and security level of the applications. Different from the authorization and authentication in Grid, this is a new privacy ensurance method enabled by sharing in the cloud.

\vspace{-1.5ex}
\subsection{Observation 5: Performance vs. Scalability}
Comparison between the implementation on HPC with implementation on the cloud is always a hot topic for scientific applications.

The NG-TEPHRA~\cite{Nunez:2010:NMP:1932688.1933014} project performed a volcanic ash dispersion simulation on both grid and cloud, using the East Cluster at Monash University and the Amazon EC2 computing resources separately. Experiments show efficient results on both platforms and the EC2 results have shown very small differences in their standard deviation, indicating the consistent QoS of the cloud. The MODISAzure~\cite{li:escience} project implemented its application on both Windows Azure cloud and a local high-end desktop machine. Evaluation on a single computational instance in Windows Azure compared with that in the desktop machine shows the task execution time of the Azure instance is always longer than that of the desktop machine while the communication time is not as stable as the computation time and does not show consistent results during the experiments. When using multiple Azure instances to compare with desktop machines, the performance of the pipeline scales almost lineally with the number of Azure instances. Cloudbursting~\cite{humphrey:assessing} implemented its satellite image processing application with three different versions: an all-cloud design on Windows Azure, a version that runs in-house on Windows HPC clusters and a hybrid cloudbursting version that utilizes both in-house and cloud resources. The hybrid version achieves the best of the previous two versions, namely the development environment of a local machine and the scalability of the cloud. Their experimental results showed that the application is benefiting from the hybrid design, both on execution time and cost.

The common observation from the above examples is that the
performance comparison between cloud and HPC is application
dependent. Due to the scheduling and communication overhead, the
applications involving large and frequent data transfer over
multiple computation nodes usually perform worse on the cloud than
on HPC clusters which are equipped with high bandwidth network. But
the advantage of cloud is its high scalability. Users can easily and
quickly scale up and down their applications as needed, without
wasting too much money. Applications such as
Cloudbursting~\cite{humphrey:assessing} can benefit from this
characteristic of the cloud.

\vspace{-1.5ex}
\subsection{Observation 6: Utilizing vs. Advancing Cloud Computing}
Many projects in various research areas are trying to benefit from
the advanced cloud computing techniques. However, most of the
eScience projects in the cloud are simply using cloud computing
techniques to improve their applications.

For example, the Climate and Earth science project SLOSH~\cite{Chandrasekar:2012:MAS:2287036.2287040}
studies the efficiency of several middleware alternatives for storm surge predictions in Windows Azure. The Life science project
CloudBLAST~\cite{Matsunaga:2008:CCM:1488725.1488913} uses MapReduce
programming model to parallelize and speedup its programs, in order
to provide distributed services for bioinformatics applications.
Another Life science project
RDF~\cite{Newman:2008:SSS:1488725.1488885} is also using MapReduce
model and Hadoop implementation to speed up the querying and
reasoning over large scale resource description framework. Many
projects in Social Sciences and Humanities are utilizing science
gateways and social networks to enable collaboration.
FMVE~\cite{Curry:2008:FMV:1437901.1438789} proposes an IT model
based on several existing technologies to enable accessing and
hosting applications on social network for enterprises.
BetterLife2.0~\cite{Hu:2010:BLS:1931470.1931831} provides
intelligent reasoning for online and mobile users through social
network interfaces. LS Gateway~\cite{10.1109/eScience.2011.16}
builds a science gateway to facilitate the sharing between life
scientists. It adopts the OAuth2.0 protocol to support authorization
for users to share data or computation resources. Many other
softwares and tools, such as the Pegasus-WMS mentioned above, are
also utilized to parallelize eScience applications and to facilitate
their execution.

A few projects dig deeper and improve the cloud computing techniques to better fit their specific applications. For example, the CGL-MapReduce project~\cite{Ekanayake:2008:MDI:1488725.1488926} proposes a new MapReduce implementation for data intensive scientific data analysis to compare with Hadoop. CGL-MapReduce uses streaming for all communications, thus eliminates the overheads in communicating via a file system.

\vspace{-1.5ex}
\subsection{Observation 7: Monetary Cost is a Concern}
Many applications have reported their implementation on the cloud
platform from the performance perspective. However, another
important consideration of eScience in the cloud, the monetary cost,
is only studied by a few example systems.

MFA~\cite{Dalman:2010:MFA:1932688.1933004} reported a 14 times
speedup for their metabolic flux analysis on Amazon cloud with a
\$11 cost, which includes the EC2 cost, EMR cost and S3 storage
cost. SciHmm~\cite{10.1109/eScience.2011.17} aims to reduce monetary
cost for scientists via deciding the most adequate scientific
analysis method for the scientists a priori. It reported the cost
for the parallel execution of SciHmm on the Amazon EC2 cloud and
showed it's acceptable for most scientists (US \$47.79). Another
project SciDim~\cite{deOliveira:2013:DVC:2465848.2465852} aims to
optimize the total execution time of scientific workflows with
budget constraints through finding the best initial configuration of
the cloud. Cloud users have to pay for all the resources they have
used on the cloud, including computation, network and storage
resources, etc. Due to the large scale of data and long running
jobs, eScience applications have to carefully plan their use of
cloud to optimize their monetary cost. However, this planning is not
trivial and requires both domain expertise and knowledge on cloud
computing.

\vspace{-1.5ex}
\subsection{Summary}
Although the current eScience system designs are far from mature,
some common trends in all of the above eScience areas have shed
light on the importance of cloud to eScience: data are easier to get
and data size is increasing tremendously; the need of sharing data
and computation and collaboration between scientists are also
increasing. Cloud computing fits in the trends perfectly. The
scalability of the cloud could offer seemingly infinite storage and
computing resources for eScience applications along with the
increase of scientific data. Also, the easy access to the cloud
resources offers great opportunity for scientists in different
locations to work on the same project.

In spite of the silver lining of developing eScience applications in
the cloud, there are still problems to solve, challenges to
overcome. The easy access to the cloud brings the security issue,
the pricing model of the cloud brings the cost-efficiency problem
and the different design between different cloud platforms also
brings us the lock-in problem. All in all, for the development of
eScience applications in the cloud, we still have a long way to go.

%% file: discussion.tex
\section{Challenges and Opportunities}\label{sec:discussion}
Previous sections have reviewed the status and the observations in
building eScience applications and systems in the cloud. Despite the
fruitful results along this research direction, we clearly see that
there are still many open problems to be addressed in order to fully
unleash the power of cloud computing for eScience. In this section,
we discuss several open problems, followed by the opportunities for
addressing those open problems.

\vspace{-1.5ex}
\subsection{Open Problems}
We present the open problems for developing the next-generation
eScience applications and systems in the cloud. Those open problems
are rooted at the interplay between eScience requirements and cloud
computing features.

\emph{ Data Lock-In:} So far, there are no eScience applications and
systems that have been deployed on multiple cloud providers. There's
no standardization between different cloud platforms, such as
different clouds use different data storage formats. For example,
data stored in Amazon S3 cannot be easily used by the jobs running
on the Windows Azure platform due to different APIs, data storage
techniques such as encryption technique and security protocols. On
the other hand, due to the eScience projects usually involve a large
amount of data for scientific research, such as the genome sequence
data and seismographic data, data transfer cost between different
cloud platforms is substantial. This also makes the data lock-in
problem significant to e-Scientists.

\emph{Performance Unpredictability:}  Some eScience applications
have rather rigid performance requirements. Performance
unpredictability is a critical problem for running those
applications in the cloud, due to the interference among concurrent
applications running in the same cloud. This problem is particularly
severe for disk I/O and network traffic. For eScience applications,
this problem is especially prominent since there are a lot of read
tasks needed to get input data and parameters from local disks to do
data analysis and also a lot of write tasks to save the intermediate
analysis results to local disks. The other factor of performance
unpredictability is VM failures or unreliability.
In~\cite{li:fault}, the authors issued a total of 10032 VM unique
instance start events on Windows Azure cloud and only 8568 instances
started once during their lifetimes while the others had encountered
various unknown problems during their run and were restarted by the
Azure infrastructure for many times.

\emph{ Data Confidentiality and Auditability:} Current commercial
clouds are essentially open to public and are consequently exposing
themselves to more attacks. Safety is the biggest concern that
prevents customers from storing their sensitive corporate data in
the cloud. Especially for eScience applications, the data involved
could be relevant to the homeland security of a country, such as the
geographical data of the country, or even the security of human
beings, such as the human genome data. So protecting these sensitive
data from unauthorized or even malicious access is an important
ongoing research topic.

\emph{Debugging:} Bugs in Large Distributed
Systems cannot be reproduced in smaller configurations. Although
many eScience programs have been tested and evaluated in the grid
and cluster environments, program debugging and testing are still
challenging in the cloud.

\emph{Lacking of eScience Common System Infrastructure.} As we
discussed in the previous section, the efforts of implementing
eScience projects on the cloud are quite ad-hoc. The effort for one
project is usually not reusable for other projects. For example, the
data processing softwares and interface APIs used in different
scientific areas are quite different. In physical sciences, the
Montage workflow, an astronomy toolkit, is commonly used to discuss
the pros and cons of using cloud computing for scientific
applications~\cite{Deelman:2008:CDS:1413370.1413421,Hoffa:2008:UCC:1488725.1488955}
and such physical science systems built in the cloud are
specifically designed to better fit the cloud for scientific
workflow applications. Thus, such developmental experiences may not
be useful to scientific applications in other areas, such as social
sciences and humanities in which resource sharing is much more the
concern and social network APIs are needed to build the social
science systems. Since current eScience systems are specifically
designed for each project, new projects coming into the cloud have
to build their systems from top down. In order to save the
development cycle and better exploit the experiences of current
systems, we need a holistic platform which applications from various
research fields can build their systems upon and offers
opportunities for application specific optimizations.

\vspace{-1.5ex}
\subsection{Opportunities}
We also see some opportunities in addressing those open problems.
Many of those opportunities are driven by different communities
outside scientists, including open-source software developers,
system researchers and governments.

\emph{Open-source Cloud Software Stacks:} With the popularity of
cloud computing, there are a lot of cloud platforms with various
architectures open to the public. Public clouds such as Amazon EC2,
Windows Azure and Google App Engine own and operate cloud
infrastructure and offer access to the public via Internet. Private
clouds implemented using software platforms such as Eucalyptus and
Nimbus on computer clusters provide hosted services to a limited
number of users behind a firewall. Private cloud is operated
for a single organization only, whether hosted and managed internally
or externally. Hybrid cloud is the composition of two or more clouds,
either private or public, to capture
the best of both worlds: ability to immediately deliver services that
users demand independent of Internet connectivity as well as the scalability to
handle cloudbursting, an instant spike in demand. Examples of hybrid
cloud include Intel Hybrid Cloud Program~\cite{intel:hybrid} and
GoGrid Cloud Hosting~\cite{gogrid}. Given an application, how to
choose the most appropriate cloud platform from the various kinds of
cloud platforms is a very challenging issue. To solve this problem,
one has to consider the characters of the application itself, such
as whether it's data-intensive or compute-intensive, whether fault
tolerance is important to this application, etc; consider the demand
of this system, such as whether the computation demand is stable or
may have instant spike of workload; also consider the aim of
sharing, for example, if the aim is to share data and resources
between limited users or the general public.

\emph{Towards Common System Infrastructure Support for eScience: }
Researchers from database community are building scientific database
which can better fit the requirements of scientific applications.
SciDB is such an example. In March 2008, the first SciDB workshop
was held in Asilomar and representatives from both scientific
community and database research community participated in this
workshop. One major result of this workshop was a set of
requirements that a database management system should meet in order
to support the storage and analysis of several fields of
data-intensive science over the next decade~\cite{BeclaLim2008}.
According to these requirements, the SciDB should provide several
new features such as direct support for arrays as a first-class
column type because all sciences need to work with non-scalar values
like vectors and arrays, association of data element with ``error
bar'' because all sciences must deal with observations and derived
data that have inherent uncertainties, etc. The SciDB developers
meeting and Open SciDB community meeting were held between 2008 to
2011 when SciDB was eventually built up and tested. The overview of
SciDB was presented at SIGMOD
2010~\cite{Brown:2010:OSL:1807167.1807271} and caught a lot of
attention from both scientific community and database community.
There have been several use cases from various sciences for SciDB
including Optical astronomy, Radio astronomy, Earth Remote Sensing,
Environmental Observation \& Modeling, Seismology and ARM Climate
Research. The aim of SciDB is to benefit all scientific applications
dealing with large-scale complex scientific analysis and provide a
way for scientists to understand data in far deeper and more natural
ways.

We have also observed many works from distributed system community
devoting to the adoption of cloud computing in scientific
environments. Youseff et al.~\cite{citeulike:5770192} establish a detailed
ontology of the cloud, dissecting the cloud into five main layers:
application layer, software environment layer, software
infrastructure layer, software kernel layer and firmware/hardware
layer. This ontology enables the scientific community to better
understand the cloud technologies and design more efficient portals and
gateways for the cloud. The Montage
Comparison example~\cite{Deelman:2008:CDS:1413370.1413421} provides
a detailed comparison between scientific workflow running in a local
environment and running in a virtual environment. The experience
shown in this paper gives the scientific community an idea what
kinds of workflows are suitable to run on the cloud and what might
be the cost if do so.~\cite{Iosup_anearly} compared the performance
of cloud to other platforms that are accessible to scientists. It
also presented two main research directions in improving the cloud
computing services for scientific computing, that is to tune
applications for virtualized resources and to optimize the
application execution considering the cost-performance-security trade-off.

To ease the pressure of scientific community, people from
distributed system community are working on simplifying the
development process of scientific applications on the cloud. Aneka
is a software platform for developing distributed
applications on private and public clouds proposed
in~\cite{Vecchiola:2009:HCC:1726593.1728946}. When implemented on
the cloud, many scientific applications need to modify their
original serial programs written in various programming models to
parallel pattern. Since Aneka supports an extensible set of
programming models, it can address a variety of different
applications and thus offers a good opportunity for scientific
applications to develop on the cloud with less effort.

\emph{National and Governmental Investment:} Another opportunity
lies in the construction of national cloud initiatives and the large
amount of funding provided by major stakeholders, such as large user
groups, vendors and governments, for cloud computing to achieve
scientific and national objectives. With the utilization of commercially
available technologies such as server virtualization,
cloud computing is able to introduce capital cost savings
to Information Technology (IT) infrastructure. Many nations have realized the
importance of cloud computing to the modernization of IT. Cloud
computing is a major feature of the US President's initiative to
modernize IT and it's also taken as an important technology for the
boost of Japan's economy by Japanese Government. Several national
cloud initiatives have been announced, including the US Cloud
Storefront, the UK G-Cloud and the Japanese Kasumigaseki.

The General Services Administration (GSA) of the US government
is an agency that focuses on implementing projects that increase
efficiencies and reduce operational cost by
optimizing common services and solutions across enterprise and
utilizing market innovations such as cloud computing services.
In September 2009, the GSA's cloud storefront Apps.gov
is launched by the Obama Administration. This online storefront
enables federal agencies to efficiently and effectively acquire
and purchase cloud computing services. Applications from desktop productivity toolsets to
document management software are now available to buy through the
online portal, which uses the software-as-a-service model to cut
government IT purchasing costs.

The G-Cloud is an iterative programme of work to achieve
government's commission to the adoption of cloud computing and
delivering computing resources, which will deliver fundamental
changes in the way the public sector procures and operates
Information and Communication Technology (ICT). At present, still in
its startup phase, the programme is resourced by existing
departmental funding allocation whilst the dedicated business case
(for 4.93 million pounds), to cover the ongoing staffing cost and
development of the CloudStore is being developed, agreed and
approved through the appropriate ministerial channels. The initial
focus of G-Cloud is on introducing cloud ICT services into
government departments, local authorities and the wider public
sector. These services can then be reviewed and purchased through
the CloudStore. At present there are 4 categories of services
provided: Infrastructure, Software, Platform and Specialist
Services. The project savings in adopting cloud computing and
re-using applications through the CloudStore can be broken down to
G-Cloud \& CloudStore and Data Centre Consolidation. It is estimated
the savings of two kinds by year 2013, 2014 and 2015 will be
\textsterling20$m$, \textsterling40$m$, \textsterling120$m$ and
\textsterling20$m$, \textsterling60$m$, \textsterling80$m$
separately.

The Kasumigaseki Cloud initiated by the Japanese Government
aims to establish a large cloud computing infrastructure to meet the
resource requirements of the Government's IT systems and enable
sharing to increase the utilization and efficiency of resources.
A National Digital Archive will also be constructed to digitize
government documents and recorded information and to improve the public access.
The concept of Green Cloud Data Centers is used to
construct the Kasumigaseki Cloud Data Center to reduce data center
energy consumption by locating them in cold regions and
increase the usage of green energy by utilizing wind
and solar power.

%% file: conclusion.tex
\section{Conclusions}\label{sec:concl}

eScience as a service is an emerging and promising service for
scientific computing. In this survey, we develop a taxonomy and
conduct a review on the current status of eScience services in the
cloud with four kinds of sciences. Compared with the relatively
mature grid infrastructure, the eScience tools and systems are in
their early stage. We believe that eScience services will be boosted
with more support from the cloud community and more investment and
efforts from the science community. We call for the combined effort
from both communities.

%% file: survey_main.bbl
\begin{thebibliography}{10}
\providecommand{\url}[1]{#1}
\csname url@samestyle\endcsname
\providecommand{\newblock}{\relax}
\providecommand{\bibinfo}[2]{#2}
\providecommand{\BIBentrySTDinterwordspacing}{\spaceskip=0pt\relax}
\providecommand{\BIBentryALTinterwordstretchfactor}{4}
\providecommand{\BIBentryALTinterwordspacing}{\spaceskip=\fontdimen2\font plus
\BIBentryALTinterwordstretchfactor\fontdimen3\font minus
  \fontdimen4\font\relax}
\providecommand{\BIBforeignlanguage}[2]{{%
\expandafter\ifx\csname l@#1\endcsname\relax
\typeout{** WARNING: IEEEtran.bst: No hyphenation pattern has been}%
\typeout{** loaded for the language `#1'. Using the pattern for}%
\typeout{** the default language instead.}%
\else
\language=\csname l@#1\endcsname
\fi
#2}}
\providecommand{\BIBdecl}{\relax}
\BIBdecl

\bibitem{A-brain-Antoniu}
G.~Antoniu, A.~Costan, B.~D. Mota, B.~Thirion, and R.~Tudoran, ``A-brain: Using
  the cloud to understand the impact of genetic variability on the brain,''
  \emph{ERCIM News}, vol. 2012, no.~89, 2012.

\bibitem{Hu:2010:BLS:1931470.1931831}
D.~H. Hu, Y.~Wang, and C.-L. Wang, ``Betterlife 2.0: Large-scale social
  intelligence reasoning on cloud,'' in \emph{CLOUDCOM '10}, 2010, pp.
  529--536.

\bibitem{Ekanayake:2008:MDI:1488725.1488926}
J.~Ekanayake, S.~Pallickara, and G.~Fox, ``Mapreduce for data intensive
  scientific analyses,'' in \emph{ESCIENCE '08}, 2008, pp. 277--284.

\bibitem{Matsunaga:2008:CCM:1488725.1488913}
A.~Matsunaga, M.~Tsugawa, and J.~Fortes, ``Cloudblast: Combining mapreduce and
  virtualization on distributed resources for bioinformatics applications,'' in
  \emph{ESCIENCE '08}, 2008, pp. 222--229.

\bibitem{GridPP:2006}
R.~W. Zurek and L.~J. Martin, ``Gridpp: Development of the uk computing grid
  for particle physics,'' \emph{Journal of Physics G: Nuclear and Particle
  Physics}, vol.~32, pp. 1 -- 20, 2006.

\bibitem{li:escience}
J.~Li, M.~Humphrey, D.~A. Agarwal, K.~R. Jackson, C.~van Ingen, and Y.~Ryu,
  ``escience in the cloud: A modis satellite data reprojection and reduction
  pipeline in the windows azure platform.'' in \emph{IPDPS'10}, 2010, pp.
  1--10.

\bibitem{li:fault}
J.~Li, M.~Humphrey, Y.-W. Cheah, Y.~Ryu, D.~A. Agarwal, K.~R. Jackson, and
  C.~van Ingen, ``Fault tolerance and scaling in e-science cloud applications:
  Observations from the continuing development of modisazure.'' in
  \emph{eScience'10}, 2010, pp. 246--253.

\bibitem{humphrey:assessing}
M.~Humphrey, Z.~Hill, C.~van Ingen, K.~R. Jackson, and Y.~Ryu, ``Assessing the
  value of cloudbursting: A case study of satellite image processing on windows
  azure.'' in \emph{eScience'11}, 2011, pp. 126--133.

\bibitem{dagman}
``Dagman: A directed acyclic graph manager,'' Condor team, July 2005,
  http://www.cs.wisc.edu/condor/dagman/.

\bibitem{Rimal:2009:TSC:1683301.1684085}
B.~P. Rimal, E.~Choi, and I.~Lumb, ``A taxonomy and survey of cloud computing
  systems,'' in \emph{NCM '09}, 2009, pp. 44--51.

\bibitem{escience:survey:web}
Association of Research Libraries, 2009,
  http://www.arl.org/rtl/eresearch/escien/esciensurvey/surveyresearch.shtml.

\bibitem{defineescience}
S.~Bohle, ``What is e-science and how should it be managed?''
  http://www.scilogs.com/scientific \textunderscore and \textunderscore medical
  \textunderscore libraries/what-is-e-science-and-how-should-it-be-managed/.

\bibitem{Globalmodelling}
US Naval Research Laboratory, Monterey, Ca.,
  http://www.nrlmry.navy.mil/sec7532.htm.

\bibitem{Newman:2008:SSS:1488725.1488885}
A.~Newman, Y.-F. Li, and J.~Hunter, ``Scalable semantics - the silver lining of
  cloud computing,'' in \emph{ESCIENCE '08}, 2008, pp. 111--118.

\bibitem{Watson08cloudcomputing}
P.~Watson, P.~Lord, F.~Gibson, P.~Periorellis, and G.~Pitsilis, ``Cloud
  computing for e-science with carmen,'' in \emph{2nd Iberian Grid
  Infrastructure Conference}, 2008, pp. 3--14.

\bibitem{top500:2010}
``Top 500 supercomputer,'' 2010, http://www.top500.org/lists/2010/11/.

\bibitem{egi}
European Grid Infrastructure, https://www.egi.eu/.

\bibitem{seti:website}
University of California, 2012, {http://setiathome.berkeley.edu/}.

\bibitem{clouddefinition}
M.~Peter and G.~Timothy, ``The nist definition of cloud computing,''
  \emph{National Institute of Standards and Technology}, October 7 2009.

\bibitem{CloudGrid360}
I.~Foster, Y.~Zhao, I.~Raicu, and S.~Lu, ``Cloud computing and grid computing
  360-degree compared,'' in \emph{GCE08}, 2008.

\bibitem{grnet:greek}
{GRNET}, http://www.grnet.gr/.

\bibitem{water:pdcc}
``Cloud-assisted real-time and large-scale monitoring and analysis for water
  quality,'' PDCC, NTU, 2011, http://pdcc.ntu.edu.sg/camawq/home.html.

\bibitem{Armbrust:2010:VCC:1721654.1721672}
M.~Armbrust, A.~Fox, R.~Griffith, A.~D. Joseph, R.~Katz, A.~Konwinski, G.~Lee,
  D.~Patterson, A.~Rabkin, I.~Stoica, and M.~Zaharia, ``A view of cloud
  computing,'' \emph{Commun. ACM}, vol.~53, no.~4, pp. 50--58, Apr. 2010.

\bibitem{Lee:2010:PSC:1851476.1851542}
C.~A. Lee, ``A perspective on scientific cloud computing,'' in \emph{HPDC '10},
  2010, pp. 451--459.

\bibitem{keahey:cloud}
K.~Keahey, ``Cloud computing for science,'' in \emph{SSDBM'09 (Keynote)}, 2009,
  pp. 1--1.

\bibitem{Oliveira2010}
D.~Oliveira, F.~Baião, and M.~Mattoso, ``Towards a taxonomy for cloud
  computing from an e-science perspective,'' in \emph{Cloud Computing:
  Principles, Systems and Applications}, 2010.

\bibitem{Hoffa:2008:UCC:1488725.1488955}
C.~Hoffa, G.~Mehta, T.~Freeman, E.~Deelman, K.~Keahey, B.~Berriman, and
  J.~Good, ``On the use of cloud computing for scientific workflows,'' in
  \emph{ESCIENCE '08}, 2008, pp. 640--645.

\bibitem{us:store}
``The us cloud storefront,'' 2009, http://www.gsa.gov/portal/content/103758.

\bibitem{uk:gcloud}
``The uk g-cloud,'' 2009,
  http://johnsuffolk.typepad.com/john-suffolk---government-cio/2009/06/government-cloud.html.

\bibitem{Kasumigaseki}
``The kasumigaseki cloud concept,'' http://www.cloudbook.net/japancloud-gov.

\bibitem{Subramanian:2010:RPS:1931470.1931899}
V.~Subramanian, L.~Wang, E.-J. Lee, and P.~Chen, ``Rapid processing of
  synthetic seismograms using windows azure cloud,'' in \emph{Proceedings of
  the 2010 IEEE Second International Conference on Cloud Computing Technology
  and Science}, ser. CLOUDCOM '10, 2010, pp. 193--200.

\bibitem{citeulike:3523379}
C.~Evangelinos and C.~N. Hill, ``Cloud computing for parallel scientific {HPC}
  applications: Feasibility of running coupled {Atmosphere-Ocean} climate
  models on amazon's {EC2},'' in \emph{Cloud Computing and Its Applications},
  2008.

\bibitem{Nunez:2010:NMP:1932688.1933014}
S.~Nunez, B.~Bethwaite, J.~Brenes, G.~Barrantes, J.~Castro, E.~Malavassi, and
  D.~Abramson, ``Ng-tephra: A massively parallel, nimrod/g-enabled volcanic
  simulation in the grid and the cloud,'' in \emph{ESCIENCE '10}, 2010, pp.
  129--136.

\bibitem{opennebula}
OpenNebula, http://opennebula.org/users:users.

\bibitem{federatedtf}
EGI Federated Cloud Task Force, https://www.egi.eu/infrastructure/cloud/.

\bibitem{gaiaspace}
GAIA-Space, http://www.esa.int/Our \textunderscore Activities/Space
  \textunderscore Science/Gaia \textunderscore overview.

\bibitem{catania}
Catania Science Gateway, http://www.catania-science-gateways.it/.

\bibitem{Deelman:2008:CDS:1413370.1413421}
E.~Deelman, G.~Singh, M.~Livny, B.~Berriman, and J.~Good, ``The cost of doing
  science on the cloud: the montage example,'' in \emph{SC '08}, 2008, pp.
  50:1--50:12.

\bibitem{Curry:2008:FMV:1437901.1438789}
R.~Curry, C.~Kiddle, N.~Markatchev, R.~Simmonds, T.~Tan, M.~Arlitt, and
  B.~Walker, ``Facebook meets the virtualized enterprise,'' in \emph{EDOC '08},
  2008, pp. 286--292.

\bibitem{Markatchev:2009:CIA:1723206.1724809}
N.~Markatchev, R.~Curry, C.~Kiddle, A.~Mirtchovski, R.~Simmonds, and T.~Tan,
  ``A cloud-based interactive application service,'' in \emph{E-SCIENCE '09},
  2009, pp. 102--109.

\bibitem{Deelman08workflowsand}
E.~Deelman, D.~Gannon, M.~Shields, and I.~Taylor, ``Workflows and e-science: An
  overview of workflow system features and capabilities,'' 2008.

\bibitem{Vockler:2011:EUC:1996109.1996114}
J.-S. V\"{o}ckler, G.~Juve, E.~Deelman, M.~Rynge, and B.~Berriman,
  ``Experiences using cloud computing for a scientific workflow application,''
  in \emph{ScienceCloud '11}, 2011, pp. 15--24.

\bibitem{Wang20121630}
J.~Wang and I.~Altintas, ``Early cloud experiences with the kepler scientific
  workflow system,'' \emph{Procedia Computer Science}, vol.~9, no.~0, pp. 1630
  -- 1634, 2012.

\bibitem{condor}
M.~Litzkow, M.~Livny, and M.~Mutka, ``{C}ondor - a hunter of idle
  workstations,'' in \emph{ICDCS}, June 1988.

\bibitem{Dean:2008:MSD:1327452.1327492}
J.~Dean and S.~Ghemawat, ``Mapreduce: simplified data processing on large
  clusters,'' \emph{Commun. ACM}, vol.~51, no.~1, pp. 107--113, 2008.

\bibitem{Ghemawat:2003:GFS:945445.945450}
S.~Ghemawat, H.~Gobioff, and S.-T. Leung, ``The google file system,'' in
  \emph{SOSP '03}, 2003, pp. 29--43.

\bibitem{Shvachko:2010:HDF:1913798.1914427}
K.~Shvachko, H.~Kuang, S.~Radia, and R.~Chansler, ``The hadoop distributed file
  system,'' in \emph{MSST '10}, 2010, pp. 1--10.

\bibitem{Ramakrishnan:2010:DFP:1807128.1807145}
L.~Ramakrishnan, K.~R. Jackson, S.~Canon, S.~Cholia, and J.~Shalf, ``Defining
  future platform requirements for e-science clouds,'' in \emph{SoCC '10},
  2010, pp. 101--106.

\bibitem{swift}
{OpenStack Swift}, https://swiftstack.com/openstack-swift/architecture/.

\bibitem{DeCandia:2007:DAH:1294261.1294281}
G.~DeCandia, D.~Hastorun, M.~Jampani, G.~Kakulapati, A.~Lakshman, A.~Pilchin,
  S.~Sivasubramanian, P.~Vosshall, and W.~Vogels, ``Dynamo: amazon's highly
  available key-value store,'' in \emph{SOSP '07}, 2007, pp. 205--220.

\bibitem{Chang:2006:BDS:1267308.1267323}
F.~Chang, J.~Dean, S.~Ghemawat, W.~C. Hsieh, D.~A. Wallach, M.~Burrows,
  T.~Chandra, A.~Fikes, and R.~E. Gruber, ``Bigtable: a distributed storage
  system for structured data,'' in \emph{OSDI '06}, 2006, pp. 15--15.

\bibitem{Lakshman:2010:CDS:1773912.1773922}
A.~Lakshman and P.~Malik, ``Cassandra: A decentralized structured storage
  system,'' \emph{SIGOPS Oper. Syst. Rev.}, vol.~44, no.~2, pp. 35--40, Apr.
  2010.

\bibitem{citeulike:8467579}
R.~Taylor, ``An overview of the {Hadoop/MapReduce/HBase} framework and its
  current applications in bioinformatics,'' \emph{BMC bioinformatics}, vol. 11
  Suppl 12, 2010.

\bibitem{Brown:2010:OSL:1807167.1807271}
P.~G. Brown, ``Overview of scidb: large scale array storage, processing and
  analysis,'' in \emph{SIGMOD '10}, 2010, pp. 963--968.

\bibitem{scidb:usecase}
SciDB Use Case, http://scidb.org/use/.

\bibitem{10.1109/eScience.2011.16}
W.~Wu, H.~Zhang, Z.~Li, and Y.~Mao, ``Creating a cloud-based life science
  gateway,'' \emph{eScience}, vol.~0, pp. 55--61, 2011.

\bibitem{Alfieri05fromgridmap-file}
R.~Alfieri, R.~Cecchini, V.~Ciaschini, and F.~Spataro, ``From gridmap-file to
  voms: managing authorization in a grid environment,'' \emph{Future Generation
  Computer Systems}, vol.~21, pp. 549--558, 2005.

\bibitem{eScience.2011.15}
A.~Nagavaram, G.~Agrawal, M.~A. Freitas, K.~H. Telu, G.~Mehta, R.~G. Mayani,
  and E.~Deelman, ``A cloud-based dynamic workflow for mass spectrometry data
  analysis,'' \emph{eScience}, pp. 47--54, 2011.

\bibitem{Thaufeeg:2011:CES:2116259.2116588}
A.~M. Thaufeeg, K.~Bubendorfer, and K.~Chard, ``Collaborative eresearch in a
  social cloud,'' in \emph{ESCIENCE '11}, 2011, pp. 224--231.

\bibitem{10.1109/TSC.2011.39}
K.~Chard, K.~Bubendorfer, S.~Caton, and O.~Rana, ``Social cloud computing: A
  vision for socially motivated resource sharing,'' \emph{TSC}, vol.~99, no.
  PrePrints.

\bibitem{Dalman:2010:MFA:1932688.1933004}
T.~Dalman, T.~Doernemann, E.~Juhnke, M.~Weitzel, M.~Smith, W.~Wiechert, K.~Noh,
  and B.~Freisleben, ``Metabolic flux analysis in the cloud,'' in
  \emph{ESCIENCE '10}, 2010, pp. 57--64.

\bibitem{conf/eScience/MudgeCHT11}
J.~C. Mudge, P.~Chandrasekhar, G.~S. Heinson, and S.~Thiel, ``Evolving
  inversion methods in geophysics with cloud computing - a case study of an
  escience collaboration.'' in \emph{eScience}, 2011, pp. 119--125.

\bibitem{Malawski:2012:CDP:2388996.2389026}
M.~Malawski, G.~Juve, E.~Deelman, and J.~Nabrzyski, ``Cost- and
  deadline-constrained provisioning for scientific workflow ensembles in iaas
  clouds,'' in \emph{SC '12}, 2012, pp. 22:1--22:11.

\bibitem{DBLP:journals/pvldb/OgasawaraOVDPM11}
E.~S. Ogasawara, D.~de~Oliveira, P.~Valduriez, J.~Dias, F.~Porto, and
  M.~Mattoso, ``An algebraic approach for data-centric scientific workflows,''
  \emph{PVLDB}, vol.~4, no.~12, pp. 1328--1339, 2011.

\bibitem{10.1109/eScience.2011.17}
K.~A. Ocana, D.~de~Oliveira, J.~Dias, E.~Ogasawara, and M.~Mattoso,
  ``Optimizing phylogenetic analysis using scihmm cloud-based scientific
  workflow,'' \emph{eScience}, vol.~0, pp. 62--69, 2011.

\bibitem{clouddrn:escience2013}
H.~Marty, S.~Jacob, I.~K. Kim, G.~K. Michael, B.~Jessica, and A.~Michael,
  ``Clouddrn: A lightweight, end-to-end system for sharing distributed research
  data in the cloud,'' in \emph{ESCIENCE '13}, 2013.

\bibitem{Chandrasekar:2012:MAS:2287036.2287040}
K.~Chandrasekar, M.~Pathirage, S.~Wijeratne, C.~Mattocks, and B.~Plale,
  ``Middleware alternatives for storm surge predictions in windows azure,'' in
  \emph{ScienceCloud '12}, 2012, pp. 3--12.

\bibitem{deOliveira:2013:DVC:2465848.2465852}
D.~de~Oliveira, V.~Viana, E.~Ogasawara, K.~Ocana, and M.~Mattoso,
  ``Dimensioning the virtual cluster for parallel scientific workflows in
  clouds,'' in \emph{Science Cloud '13}, 2013, pp. 5--12.

\bibitem{intel:hybrid}
Intel Hybrid Cloud Program, http://www.intelhybridcloud.com/.

\bibitem{gogrid}
The GoGrid Cloud Hosting, http://www.gogrid.com/cloud-hosting/case-studies/.

\bibitem{BeclaLim2008}
J.~Becla and K.-T. Lim, ``Report from the scidb workshop,'' \emph{Data Science
  Journal}, vol.~7, no. September, pp. 88--95, 2008.

\bibitem{citeulike:5770192}
L.~Youseff, M.~Butrico, and D.~Da~Silva, ``Toward a unified ontology of cloud
  computing,'' in \emph{GCE '08}, 2008, pp. 1--10.

\bibitem{Iosup_anearly}
R.~Iosup, S.~Ostermann, N.~Yigitbasi, R.~Prodan, T.~Fahringer, and D.~Epema,
  ``An early performance analysis of cloud computing services for scientific
  computing,'' \emph{TU Delft, Tech. Rep., Dec 2008, [Online] Available}, 2008.

\bibitem{Vecchiola:2009:HCC:1726593.1728946}
C.~Vecchiola, S.~Pandey, and R.~Buyya, ``High-performance cloud computing: A
  view of scientific applications,'' in \emph{ISPAN '09}, 2009, pp. 4--16.

\end{thebibliography}
